\documentclass[usedcolumn,usenatbib,usegraphicx]{mn2e}
\usepackage{lscape}
\usepackage{setspace}
\title[Studies of Quasi-Periodic Oscillations in the Black Hole Transient XTE J1817-330]
  {Studies of Quasi-Periodic Oscillations in the Black Hole Transient XTE J1817-330}
\author[J.Roy, P.C.Agrawal, B.Paul, K.Duorah]
  {Jayashree~Roy$^{1,3}$, \thanks{E-mail: jayashree@tifr.res.in}
   P.C. Agrawal$^1$, B.Paul$^2$ and K.Duorah$^3$ \\
  $^1$Tata Institute of Fundamental Research, Mumbai 400 005, India\\
  $^2$Raman Research Institute, Bangalore, 560080 , India \\
  $^3$Gauhati University, Guwahati 781014, India}
\date{Received ...,; accepted ...}

\pagerange{\pageref{firstpage}--\pageref{lastpage}} \pubyear{2009}

\def\LaTeX{L\kern-.36em\raise.3ex\hbox{a}\kern-.15em
    T\kern-.1667em\lower.7ex\hbox{E}\kern-.125emX}

\begin{document}

\label{firstpage}

\maketitle
\begin{abstract}
We have used archival RXTE PCA data to investigate timing and spectral characteristics of the transient XTE J1817-330. The data pertains to 160 PCA pointed observations made during the outburst period 2006, January 27 to August 2. A detailed analysis of Quasi-Periodic Oscillations (QPOs) in this black hole X-ray binary is carried out. Power density spectra were obtained using the light curves of the source.
QPOs have been detected in the 2-8 keV band in 10 of the observations. In 8 of these observations, QPOs are present in the 8-14 keV and in 5 observations in the 15-25 keV band. XTE J1817-330 is the third black hole source from which the low frequency QPOs are clearly detected in hard X-rays. The QPO frequency lies in $\approx$ 4-9 Hz and the rms amplitude in 1.7-13.3\% range, the amplitude being higher at higher energies. We have fitted the PDS of the observations with Lorentzian and power law models. Energy spectra are derived for those observations in which the QPOs are detected to investigate any dependence of the QPO characteristic on the spectral parameters. These spectra are well fitted with a two component model that includes the disk black body component and a power law component. The QPO characteristics and their variations are discussed and its implication on the origin of the QPOs are examined. 
     
\end{abstract}

\begin{keywords}
 stars: black hole binary: individual: XTE J1817-330 stars
\end{keywords}

\section{Introduction}
\label{sec:intro}

Accretion powered X-ray binaries are the brightest X-ray sources in our Galaxy. These binaries contain either an accreting neutron star or an accreting black hole as the X-ray source. Based on estimates of the mass of the accreting object and its X-ray characteristics, about 40 X-ray sources have been classified as black hole binaries (Remillard and McClintock 2006a). Of these 20 have reliable mass estimates and are, therefore, regarded as confirmed black holes while the remaining 20 are considered to be black hole candidates (Remillard and McClintock 2006a). A majority of the black hole binaries (BHBs) are transients and most of them have a low mass optical companion.\\

The black hole X-ray binaries have several distinctive X-ray characteristics. During the outburst there is a strong soft component that originates in the inner region of the hot accretion disk. The presence of a hard X-ray component in energy spectra is another common feature of black hole binaries. The hard X-rays arise through Compton scattering of low energy photons from the accretion disk in a hot optically thin plasma. The resulting hard X-rays are referred to as the thermal Compton component (McClintock and Remillard 2006).\\

Low frequency quasi-periodic oscillations (QPOs) in $\approx$ 0.1-40 Hz range and high frequency QPOs $\approx$ 50-450 Hz also occur in many BHBs. Presence of several distinct spectral states and transition from one state to another at irregular intervals, is another distinguishing feature of BHBs. Spectral and temporal characteristics of the sources vary from one state to another. Remillard and McClintok (2006a) have broadly classified the spectral state for BHB as (a) High or Soft state (HS) dominated by the thermal component, (b) Low or Hard State (LH) marked by the hard X-ray power law and (c) Steep power law (SPL) or very high state characterized by steep slope power law component. For a more complete description of the spectral states McClintock and Remillard (2006) included two more states namely an Intermediate state (IM) that occurs when the source moves from LH to HS state and an extreme LH state which they called as quiescent state. Gierlinski, Done and Page (2008) have included an additional state termed as Ultra soft state (US) which is an extreme case of high/soft state found in this source. The US state is characterized by a very weak high energy tail with a low disk temperature and very low hardness ratio $\leq$ 0.1 (Gierlinski, Done and Page 2008). The low frequency QPOs are usually detected in the SPL and LH states and have been observed in 14 BHBs so far (Remillard and McClintock 2006a). Detection of LFQPO in XTE J1817-330 indicates that LFQPOs can also be found sometime in the HS state.\\

An X-ray transient known as XTE J1817-330 was discovered by Remillard et al. (2006b) on 2006 January 26, with the All Sky Monitor (ASM) on Rossi Timing X-ray Explorer (RXTE) (Levine et al. 1996). At the time of its detection the 2-12 keV flux was  0.93 Crab. The intensity then rose to a peak value of 1.9 Crab on January 28, and then declined to 1.2 Crab by January 30. Subsequently it decayed exponentially with a decay time of 27 days (Sala et al. 2007). From its high/soft state at the time of the outburst, the source declined to a low/hard state characterized by kT = 0.2 keV. A change in the intensity state of the source occurred around $\sim$ February 9 (Shaw et al. 2006). Hard X-rays (20-60 keV) flux increased from 46 $\pm$ 2 mCrab on February 9  to 79 $\pm$ 2 mCrab on February 14 accompanied by a decrease of soft X-ray (2-10 keV) flux from 840 $\pm$ 4 mCrab to 670 $\pm$ 4 mCrab during this period indicating the onset of the hard state (Kuulkers et al. 2006).\\

Its energy spectrum was studied with the RXTE, XMM-Newton, Integral and Swift instruments. Sala et al. (2007) measured the spectra using data from the XMM-Newton and Integral instrument when the source was in a high/soft state during 2006 February-March. Their spectral results indicated that its energy spectrum was typical of a BHB source with a dominant thermal disk component well described by kT $\sim$ 0.7-0.9 keV and a thermal Compton power law component with a photon index of $\sim$ 2-3 (Sala et al. 2007). Its spectral characteristics were observed with X-ray telescope (XRT) on the Swift satellite covering different stages of the outburst over 160 days. During this period the source made transition from the high/soft state in the initial outburst to a low hard state near the end of the outburst. The XRT spectra in the high soft state in 0.6-10 keV are well described by a two component model consisting of a thermal component from a optically thick and geometrically thin accretion disk and a hard power law component. The temperature of the inner disk producing the soft component declined from $\sim$ 0.8 keV during the initial outburst phase to $\sim$ 0.2 keV near the end of the outburst (Rykoff et al. 2007).

 A detailed study of the spectral evolution of XTE J1817-330 at different phases of the outburst was carried out by Gierlinski, Done and Page (2008) using the RXTE and the Swift data. The spectra of 150 PCU2 (RXTE) observations were modeled with the two component spectral model consisting of a disk component and a hard thermal Compton power law component. Using the same XRT data as used by Rykoff et al. (2007), they also found that the inner disc temperature declined from $\sim$ 0.9 keV to $\sim$ 0.2 keV as the source intensity declined. Based on this they claimed that the accretion disk recedes when the source transits from the high/soft state to the low/hard state. It may also be noted that apart from XTE J1118+480, this black hole binary has the lowest absorption along the line of sight among all the bright black hole candidates as obtained from the Chandra and Swift spectral data (Miller et al. 2006a;b).  \\

Power density spectra of XTE J1817-330 obtained from the first two X-ray observations during 06:03-17:04 UTC on 2006 February 24, revealed strong QPOs at $\sim$8.5 Hz (Homan, Miller and Wijnands 2006). Third observation (13:20-13:43 UTC) on the same day showed only a weak QPO at around 6.4 Hz. The last 2 observations (14:55-17:04 UTC) again indicated the presence of strong QPOs at 5.0 Hz.\\

Rupen, Dhawan and Mioduszewski (2006a), detected with VLA a radio object in the error box of the X-ray source having a flux density of 2.1 mJy at 1.4 GHz on 2006 January 31. The radio  source faded away by 2006 February 2 (Rupen, Dhawan and Mioduszewski 2006b). A bright optical counterpart of the X-ray source was found at the time of the outburst with V = 11.3 magnitude and its brightness decreased to V = 15.5 by February 10 (Torres et al. 2006). The optical star was also detected in the near-infrared with a K magnitude of 15.0 on 2006 February 7 (D'Avanzo et al. 2006). Near-UV observations with the Optical Monitor on the XMM-Newton showed variations in the UV flux correlated to the hard X-ray flux variations (Sala et al. 2007).  \\

All these characteristics strongly suggest that XTE J1817-330 is most likely a black hole binary. This transient was repeatedly observed with the Proportional Counter Array (PCA) on RXTE in the pointed mode during the period 2006 January 27 - August 2. We have carried out detailed timing analysis of the PCA data to study the properties of the QPOs and in this paper we present the results of this analysis.

\section{Observations and Data Reduction}
\label{observation section}

The data for our analysis are taken from the HEASARC data archive (http://heasarc.gsfc.nasa.gov/). We have used data acquired from the observations obtained during 2006, January 27 - August 2 with the 160 PCA pointing's.
The PCA consists of five xenon proportional counter units (PCUs) sensitive in 2-60 keV energy range with a total effective area of $\sim$6500 cm$^{2}$ at $\sim$ 10 keV (Jahoda et al. 1996). We used data from only the PCU2 for generating the light curves, energy spectra and the hardness ratio of the source as this unit was operating in all the observations.\\

FTOOLS version 6.1 was employed for the analysis and the calibration data files of epoch 3 and 4 were used for the energy response matrix. The XTE filter file is created using task xtefilt in FTOOLS with a time step of 16 second. The binned mode data were reduced to create the light curve files for all the observation Id's using saextrct with a bin size of 7.8125 milli second. Event mode data were used for constructing the power density spectra (PDS) in the 15-25 keV energy range. These data were extracted from the light curve files using the same binning time as used for the binned mode data using seextrct task in FTOOLS. The spectral studies of XTE J1817-330 were carried out by using standard 2 mode data for the PCU 2 with a binning time of 16 sec.

\section{Data Analysis and Results}
\label{Analysis section and Results}
\begin{figure}
\resizebox{\hsize}{!}{\includegraphics*[width=8cm, angle=-90]{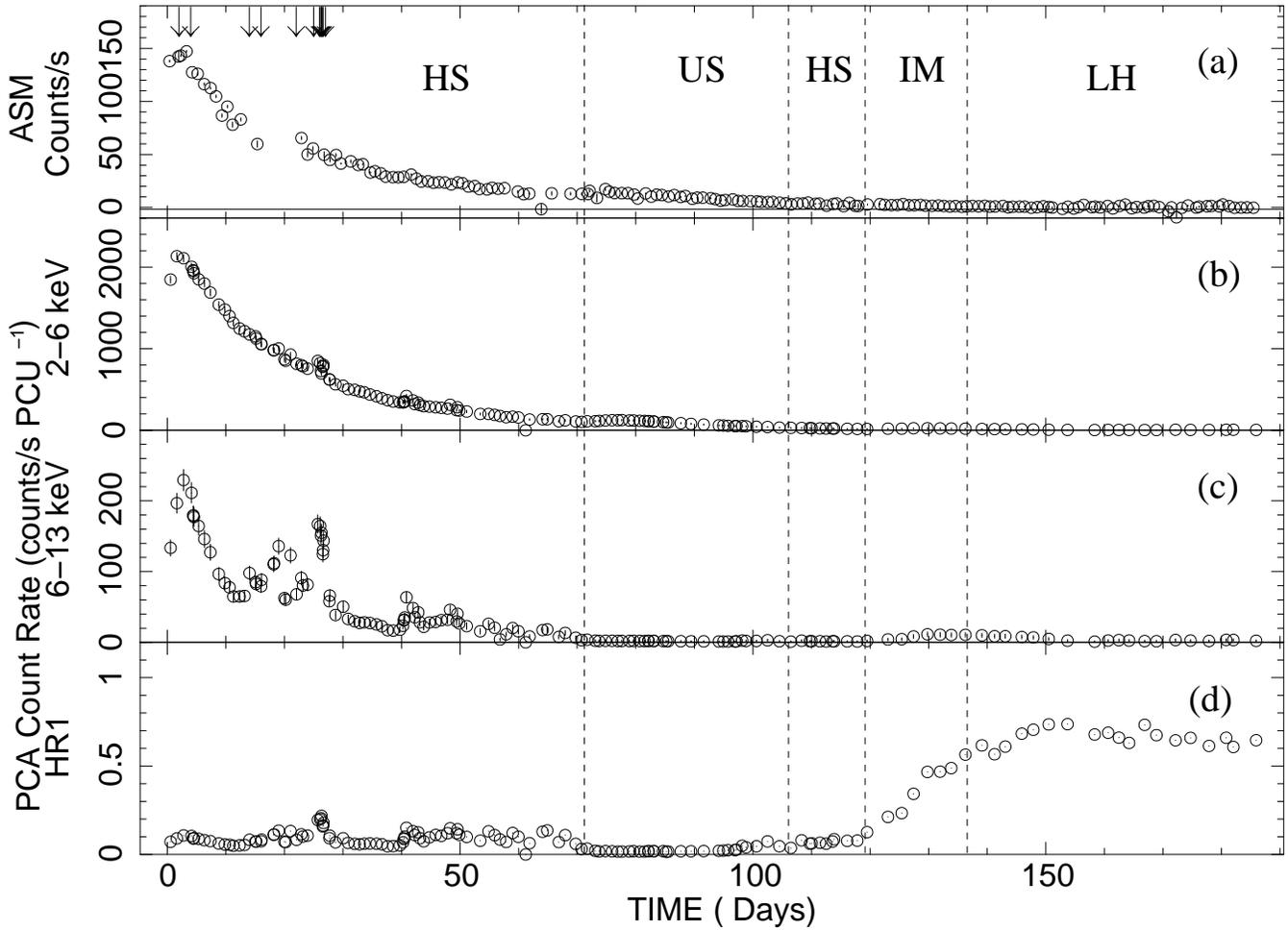}}
\caption{The ASM lightcurve in the 1.5 -12 keV range is shown in panel (a) from 2006-Jan-29 (MJD 53764) to 2006-Aug-02 (MJD 53950). The count rates are averaged over a day. The position of the observed QPOs are indicated by the vertical arrows. The PCA light curves in the 2-6 keV and 6-13 keV energy bands are shown in (b) and (c). The corresponding hardness ratio obtained from counts in 6-13 keV / 2-6 keV is shown in panel (d) for the 100 pointed PCA observations.}
\label{fig:fig1}
\end{figure}

\begin{figure}
\resizebox{\hsize}{!}{\includegraphics*[width=8cm, angle=-90]{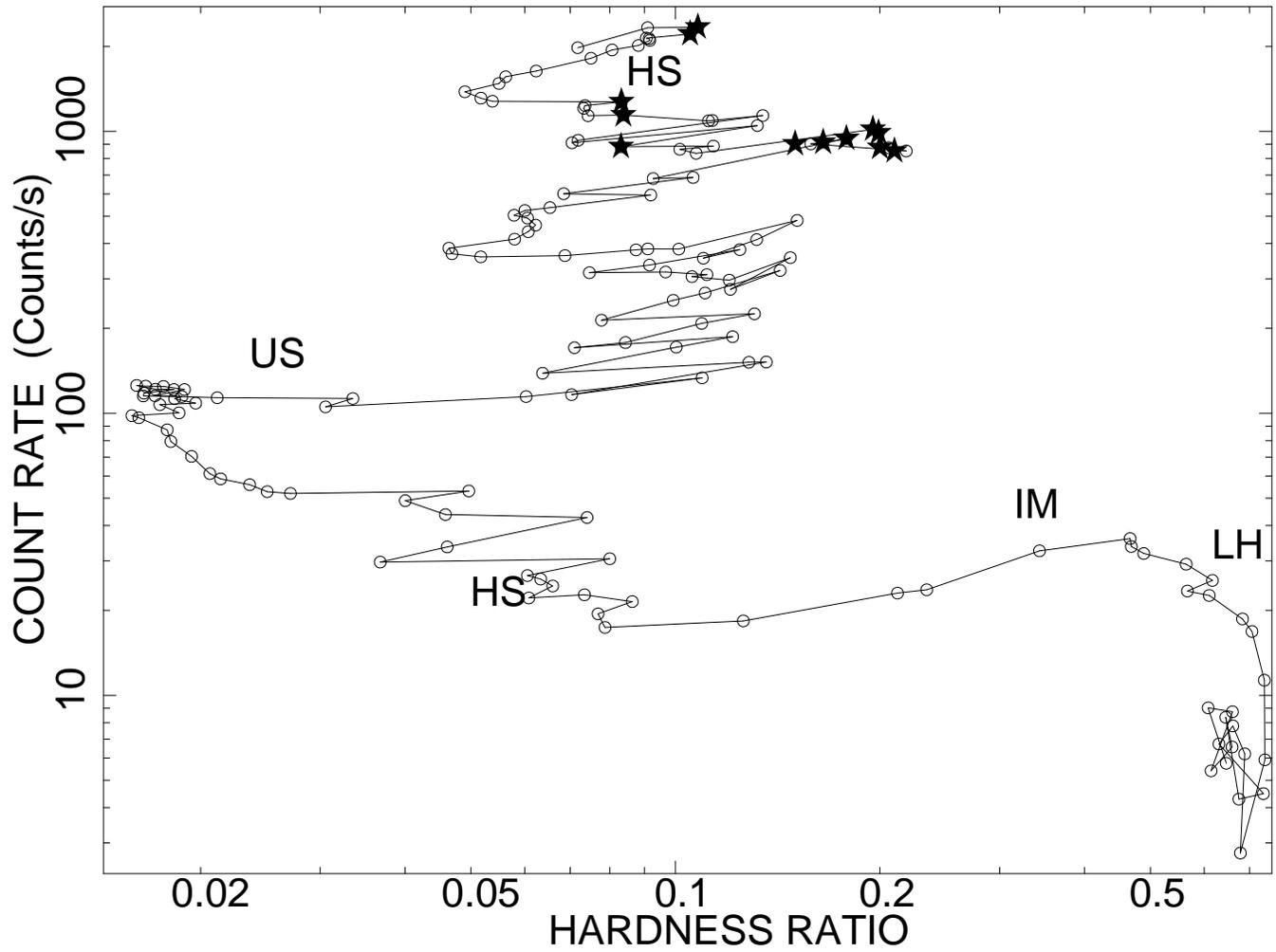}}
\caption{Variation of the Hardness Ratio (counts in (6-13) keV / counts in (2-6) keV) with the count rate (s$^{-1}$) in (2-13) keV from only PCU 2, is shown during the 2006 outburst of the source. Filled $\star$ symbols represent the positions where the QPOs are detected.}
\label{fig:fig2}
\end{figure}

The X-ray light curve of the transient was constructed by using the 1.5-12 keV count rates from the ASM and it is shown in Fig \ref{fig:fig1}. In the ASM light curve the source reached a peak intensity of about 1.9 Crab on 2006 January 28 and then declined with an e-folding time of about 27 days. There is indication of a broad peak between 2006 January 28 to 2006 February 2. After a lapse of about 120 days since the peak intensity, the source became undetectable. Similar light curves were obtained by Rykoff et al. (2007) and Sala et al. (2007) using the ASM daily average count rates for the entire duration of the outburst. Sala et al. (2007) also derived variation of the hardness ratios using the count rates in 3.0-5.0 keV / 1.5-3.0 keV and 5.0-12 keV / 3.0-5.0 keV using the ASM data from 2006 January 30 to April 30. We have computed the hardness ratio [(6-13) keV counts / (2-6) keV counts] from the PCU 2 data and this is shown as a function of the source intensity in Fig \ref{fig:fig2}. We used the spectral state classification of Gierlinski, Done and Page (2008) to indicate the spectral state of the source in Figures \ref{fig:fig1} and \ref{fig:fig2}.
From MJD 53764 (2006 January 29) the source was in HS state for 70 days with a hardness ratio less then 0.2. The source then moved to the US state characterized by a very low hardness ratio ($\ll$ 0.1), and again appeared to move back to HS state. After 120 days of the HS state, the source, passed through 15 days of intermediate state with hardness ratio in the range 0.2-0.4. Finally it reached the LH state with a hardness ratio $>$ 0.4. Gierlinski, Page and Done (2008) had also presented a similar plot (Figure \ref{fig:fig2}) of hardness ratio obtained from 6.3-10.5 keV count rates / 3.8-6.3 count rates using the PCA (RXTE) data. There is close resemblance between the two sets of curves even though the energy bands are different.\\ 

\subsection{QPO Analysis and Results}
\label{QPO Analysis and Results}

The PDS is a powerful method for probing the rapid variability in the black hole and other accretion powered X-ray binaries. The PDS of many BHBs exhibits narrow and broad QPOs peaks whose width and position vary with time.
A search for the QPOs in XTE J1817-330 was carried out in 2-8 keV, 8-14 keV and 15-25 keV bands. The power density spectra were constructed for all the observations using powspec. All the power spectra were normalized such that their integral gives the squared rms fractional variability (therefore the power spectrum is in units of (rms)$^{2}$/Hz) with the expected white noise level subtracted. The binned mode data were used for the 2-8 keV and 8-14 keV bands and the event mode data were used for the 15-25 keV for generating the PDS. The power law and the Lorentzian models have been used for fitting the QPO profiles. A QPO feature is detected in the 2-8 keV band in 10 of the observations. The observations showing presence of the QPOs are indicated by vertical arrows in the ASM light curve in Fig \ref{fig:fig1}. It may be noted that the QPOs are detected only when the source was bright (40-150 ASM counts sec$^{-1}$). As the source intensity declined to a level below 40 ASM counts sec$^{-1}$, the QPOs disappeared. The QPO occurrence is also indicated in the plot of hardness ratio versus source intensity in Fig \ref{fig:fig2} by star symbol. In 6 of these 10 observations, the QPOs are present in the 8-14 keV energy band. At the higher energy (15-25 keV), the QPOs are detected only in 5 of the observations. \\

\begin{figure}
\resizebox{\hsize}{!}{\includegraphics*[width=10cm, height=8cm, angle=-90]{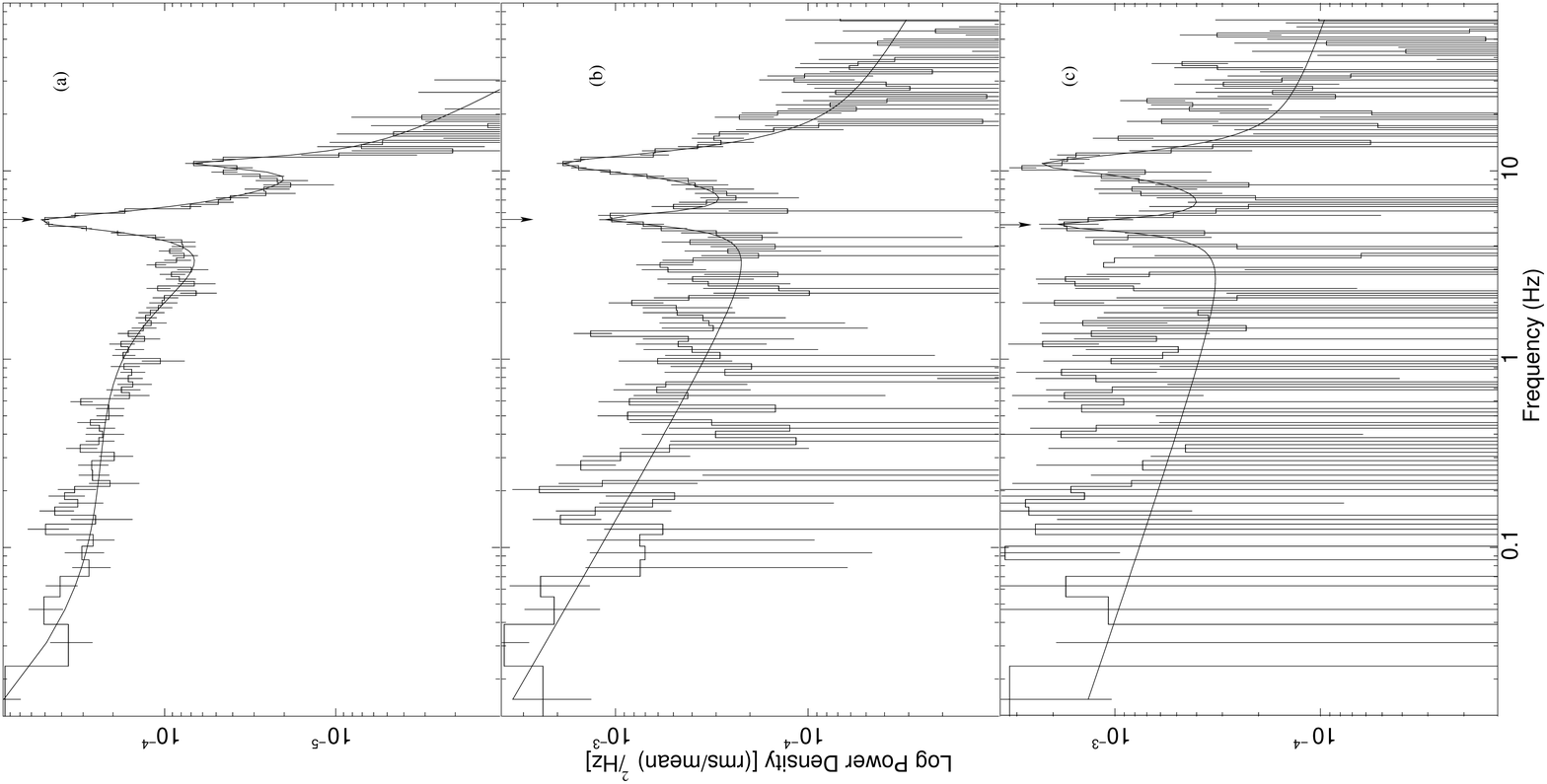}}
\caption{Power density spectra for the observation of MJD 53768. The PDS is shown for 2-8 keV in panel (a), 8-14 keV in panel (b) and 
15-25 keV in panel (c). Arrows in the panels (a), (b) and (c) indicate the fundamental frequencies of the QPOs at 5.39 Hz, 5.55 Hz and 5.14 Hz. The first harmonic at $\approx$ 10 Hz is more prominent than the fundamental frequency peak in panel (b) and (c).} 
\label{fig:fig3}
\end{figure}

\begin{figure}
\resizebox{\hsize}{!}{\includegraphics*[width=10cm, height=8cm, angle=-90]{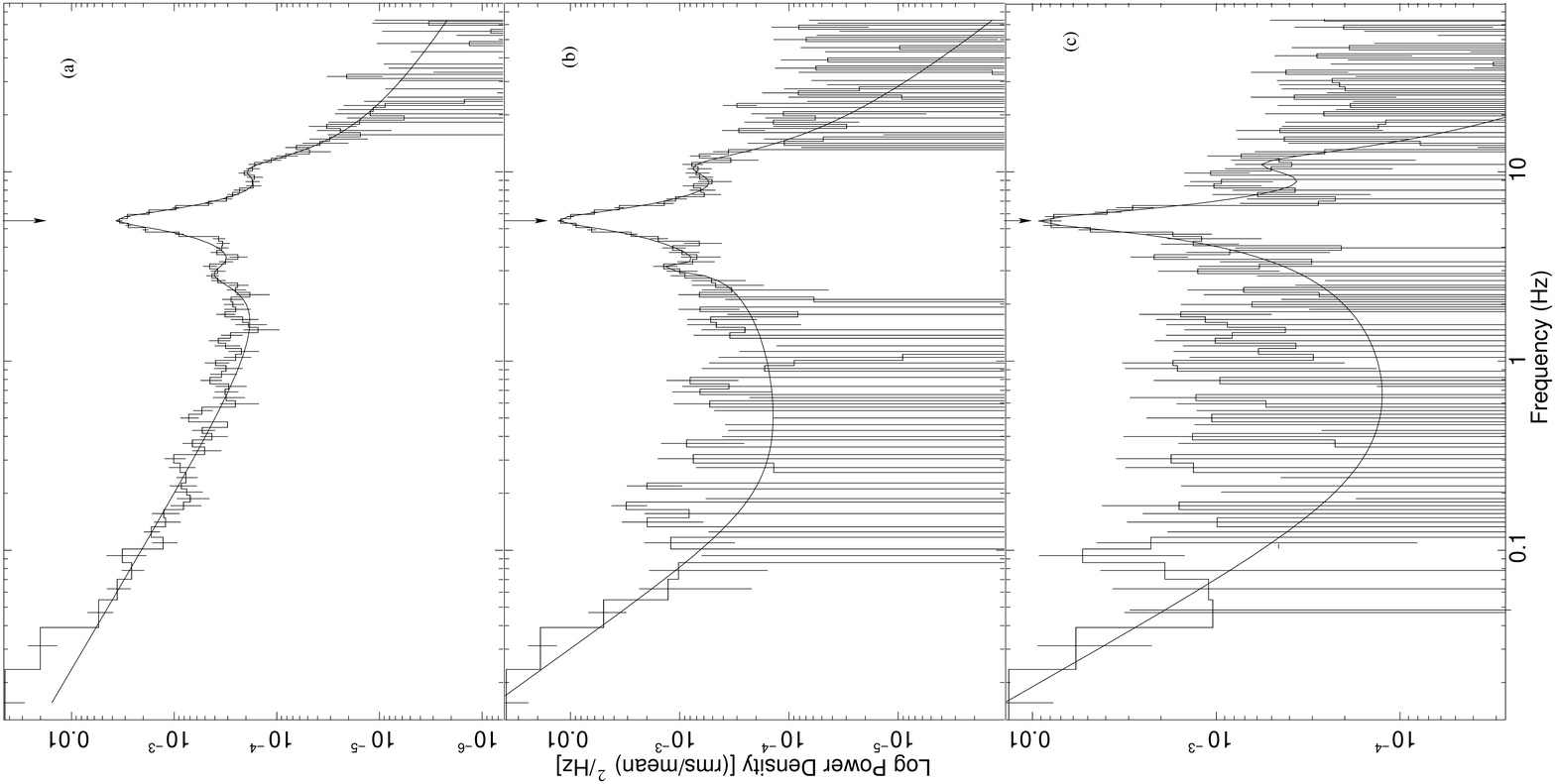}}
\caption{Power density spectra for the observations of MJD 53789. The PDS is shown for 2-8 keV in panel (a), 8-14 keV in panel (b) and 
15-25 keV in panel (c). Arrows in the panels indicate the fundamental frequencies of the QPOs at 5.57 Hz, 5.57 Hz and 5.5 Hz. The first harmonic at $\approx$ 10 Hz is weaker compared to the fundamental frequency at $\approx$ 5 Hz in all the panels.} 
\label{fig:fig4}
\end{figure}

\begin{figure}
\resizebox{\hsize}{!}{\includegraphics*[width=10cm, height=8cm, angle=-90]{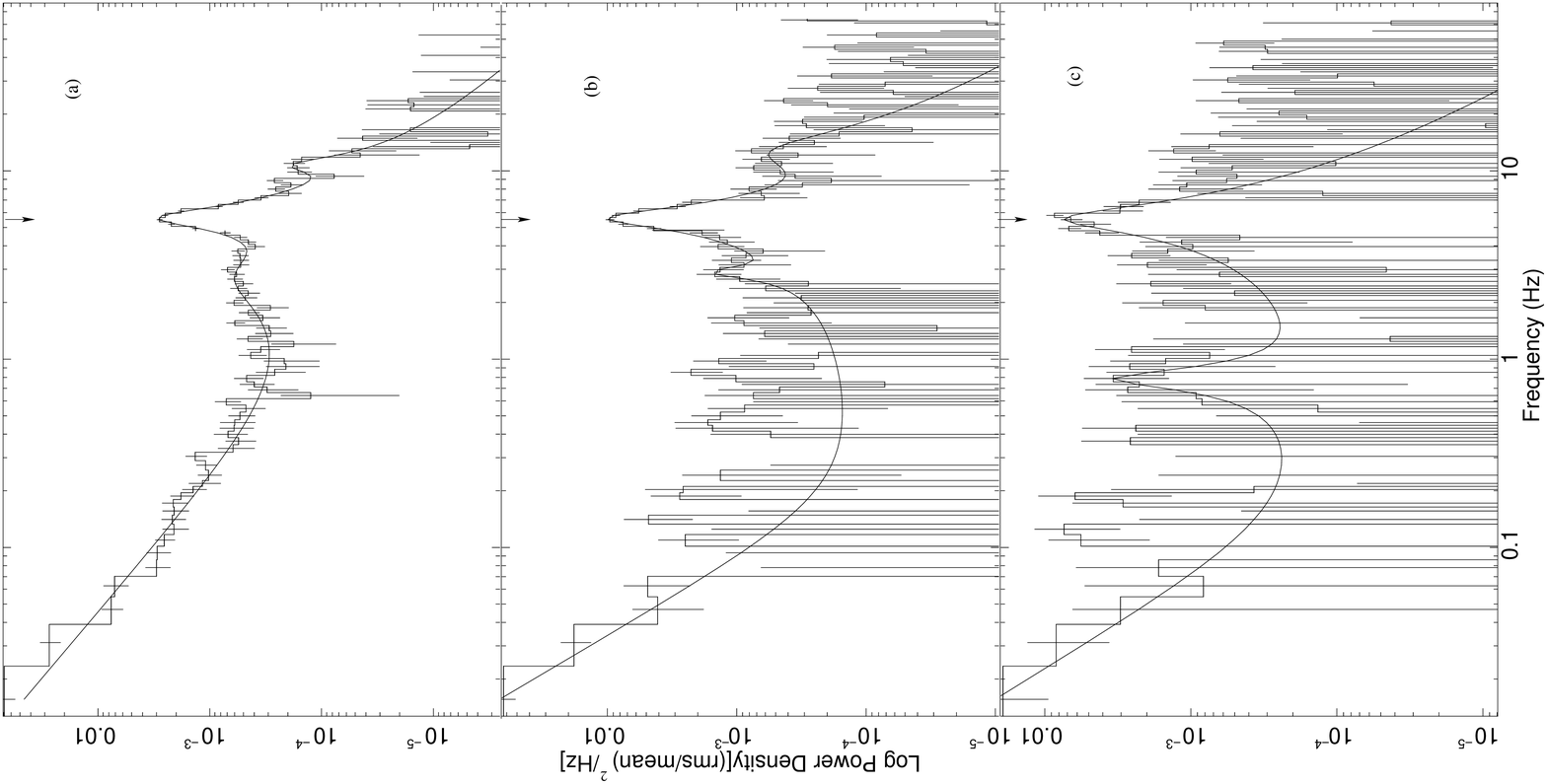}}
\caption{Power density spectra for observation of MJD 53790. The PDS is shown for 2-8 keV in panel (a), 8-14 keV in panel (b) and 
15-25 keV in panel (c). Arrows in the panel (a), (b) and (c) indicate the fundamental frequencies of QPOs at 5.61 Hz, 5.60 Hz and 5.54 Hz.}
\label{fig:fig5}
\end{figure}

\begin{figure}
\resizebox{\hsize}{!}{\includegraphics*[width=10cm, height=8cm, angle=-90]{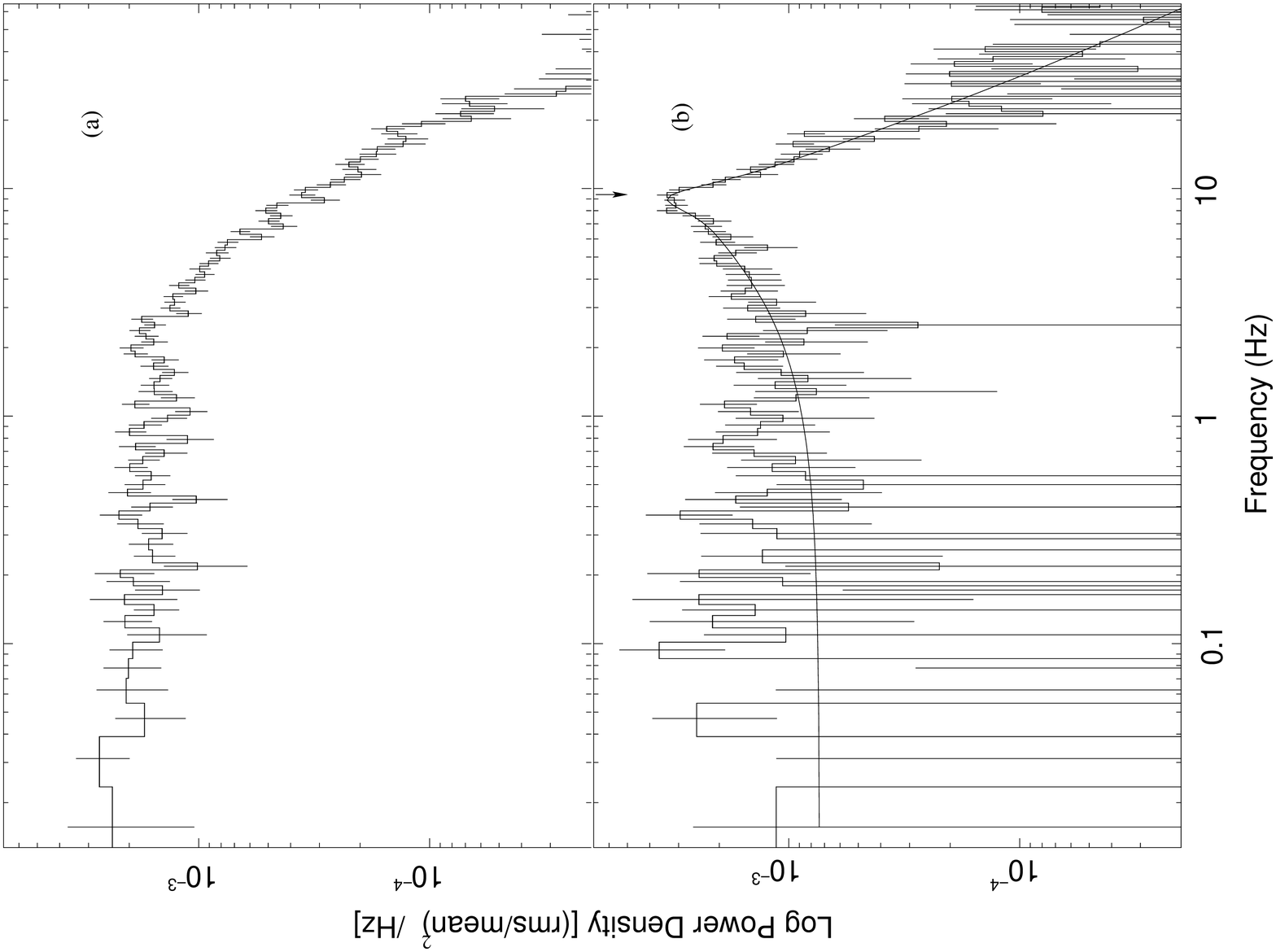}}
\caption{Power density spectrum for the MJD 53790.2 is shown in panel (a)for 2-8 keV, and (b)for 8-14 keV energy band. The arrow indicate the position of broad QPO at 9.06 Hz in panel(b). The QPOs are not detectable in (2-8) keV but a broad and asymmetric peak is present in panel (b) at about 9 Hz.}
\label{fig:fig6}
\end{figure}

The power density spectra in the different energy ranges for three of the observations are shown in Figures \ref{fig:fig3}, \ref{fig:fig4} and \ref{fig:fig5}. Prominent QPO peaks are present in the PDS. The QPO feature at $\approx$ 5 Hz is detected prominently in the higher energy band (15-25 keV). A first harmonic of the fundamental QPOs at $\approx$ 10 Hz is also present in the PDS. Note that the first harmonic is quite prominent in the 2-8 keV (Fig \ref{fig:fig3}a) and the peak at $\approx$ 10 Hz in Fig \ref{fig:fig3}(b) for 8-14 keV band is even stronger then the peak at $\approx$ 5 Hz. The fundamental QPO peaks appear at about the same frequency in the plots in the three different energy bands. 
The power density spectra for MJD 53790.2 are shown in Fig \ref{fig:fig6}. There is no indication of the presence of QPOs in 2-8 keV but a broad QPO peak is clearly seen at about 9 Hz in the PDS of 8-14 keV. It is conceivable that the 5 Hz peak is blended in the rather broad 9 Hz peak. Note that if the 9 Hz peak is identified as the first harmonic as seen in Figs 3, 4 and 5, the peak due to fundamental QPO frequency at $\approx$ 5 Hz is undetectable in Fig 6(a) and (b). Small but insignificant peaks can be seen in Fig \ref{fig:fig4} at $\sim$ 3 Hz in 2-8 keV and 8-14 keV bands and at $\sim$ 3 Hz and $\sim$ 0.8 Hz in 8-14 keV and 15-25 keV bands in Fig \ref{fig:fig5}.\\

A summary of the characteristic of the QPOs eg., frequency, amplitude and width for all the observations that showed the presence of the QPOs, is presented in Table \ref{table1.tab}. It may be noticed from the table that the data of MJD 53790.2 and 53790.3, show no detectable QPOs in 2-8 keV. However the QPOs are clearly present in the 8-14 keV at a higher frequency of $\approx$ 8-9 Hz. The QPO peaks in the 8-14 and 15-25 keV bands are rather broad with $\delta$$\nu$ in 1.8 to 3 Hz range. This is unlike the QPOs detected in the other observations where the peaks are narrow with $\delta$$\nu$ less then 1 Hz. Consequently the Q values are rather low being 3 to 4. 
The values of the coherence parameter (quality factor, Q$=$$\nu$/$\delta\nu $) are all greater than 2.\\

We found no correlation in the amplitude of fundamental QPO and amplitude of first harmonics. The QPO amplitudes were not found to show any correlation with the overall source intensity or hardness ratio. No trend was observed between the QPO amplitudes derived in the two energy bands 2-8 keV and 8-14 keV. Similarly no correlation was detected in the variation of power-law flux with QPO fundamental frequency and rms amplitude of the QPOs. 

We found from the PDS studies on 2006 February 24 that the QPOs were present in high/soft state of the source with spectral power-law index varying in the range 2.1-2.3. Our result are supported by the findings of Homan et al. (2006) who discovered rapid variability in the QPO properties on the same day. The QPOs were observed in high/soft state of the source with the spectral power-law index from $\sim$ 2.3-2.4 (Homan et al. 2006). 

\begin{figure}
\resizebox{\hsize}{!}{\includegraphics*[width=8cm, angle=-90]{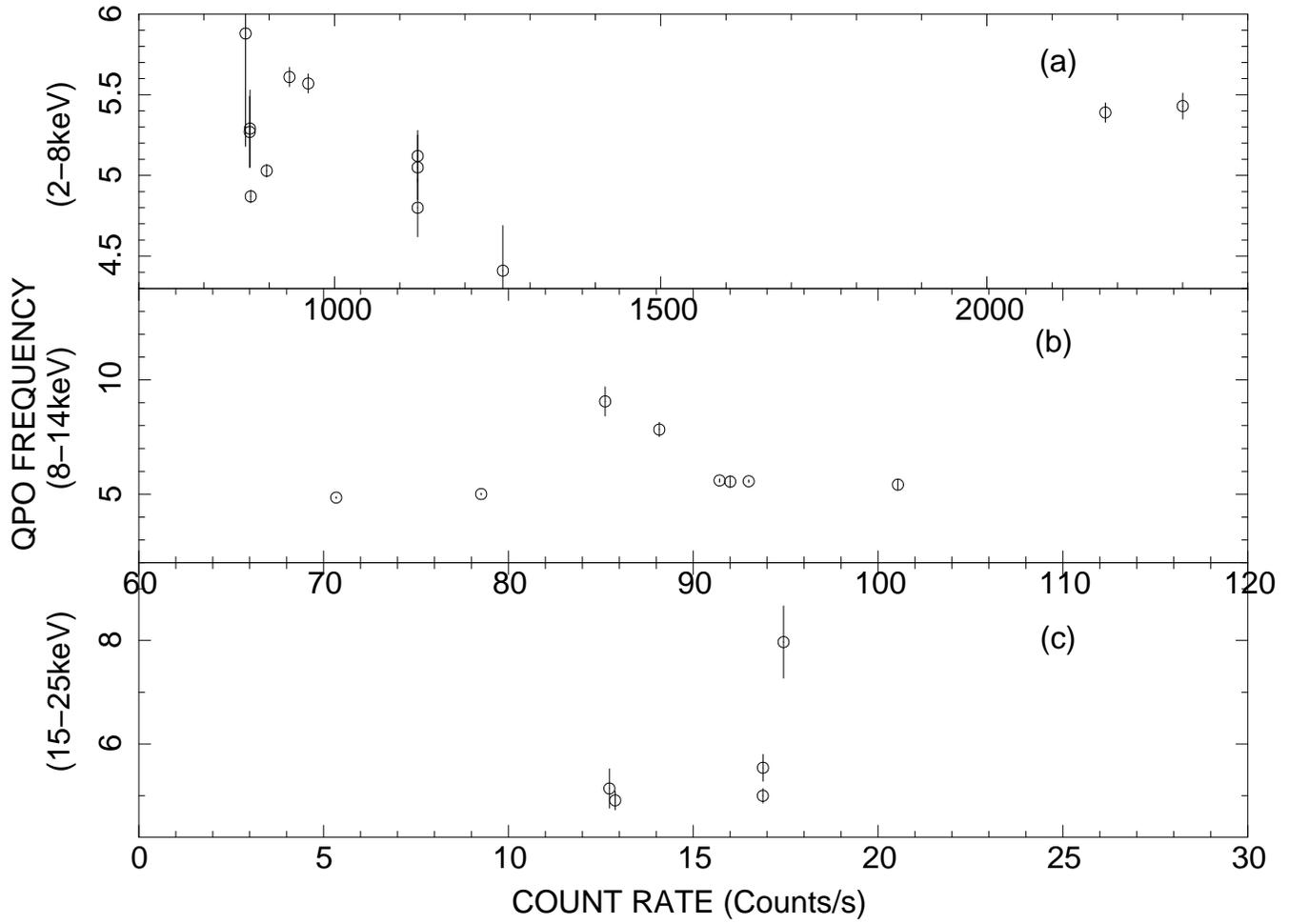}}
\caption{Plots of the QPO frequency versus the source intensity are shown in the three energy bands. These are background subtracted source count rates (counts s$^{-1}$) taken only from  PCU 2 in the energy intervals (a) 2-8 keV, (b)8-14 keV and (c)15-25 keV.}
\label{fig:fig7}
\end{figure}

To investigate whether the QPO frequency has any dependence on the source intensity, we have plotted the QPO frequency versus the source count rate in the three energy bands in Fig \ref{fig:fig7}. These are background subtracted counts per second taken only from the PCU2 that was working in all the observations. There is indication of decreasing trend in the QPO frequency from $\approx$ 5 Hz to $\approx$ 4 Hz as the source intensity is increased [Fig \ref{fig:fig7}(a)]. The frequency, however, again increased as the source intensity increased by a factor of more than two. In the 8-14 keV and 15-25 keV channels no correlation is obvious between the frequency and the intensity in Figures \ref{fig:fig7}(b), \ref{fig:fig7}(c). 

Hence it may be inferred that there is no clear trend of a change of the QPO frequency with the intensity. The QPO frequency varies in an erratic manner in a narrow band of 4.4 Hz to 5.9 Hz in the 2-8 keV channel.\\

\subsection{Spectral Analysis and Results}
\label{Spectral Analysis and Results}

We have analyzed the spectral data from the RXTE for those observations in which the QPOs are detected. For a comparison the spectra are also obtained for a few observations in which no QPOs are detected to investigate whether there are any differences in the spectral parameters. Background subtracted standard 2 mode data from the PCU 2 with 16 sec binning were used to construct the spectra. The energy spectra of selected observations were fitted with a power law model taken from XSPEC version 12 for high energy component of the spectrum, plus a standard disk black body, diskbb model (accretion disk consisting of multiple blackbody components) taken from XSPEC (Mitsuda et al. 1984 ; Makishima et al. 1986). It also included the photoelectric absorption (wabs) model from (Morrison and McCammon, 1983) and the Xenon edge at $\approx$ 4.7 keV to account for the PCA response and is not intrinsic part of the spectrum. In the spectrum of MJD 53768 a Gaussian line model is also added to account for the presence of an iron line at 6.4 keV. In the spectral fits a fixed value of hydrogen column density (N$_{H}$)= 1.2$\times$10$^{21}$ cm$^{-2}$ has been used (Rykoff et al. 2007). \\

\begin{figure}
\resizebox{\hsize}{!}{\includegraphics*[width=8cm, angle=-90]{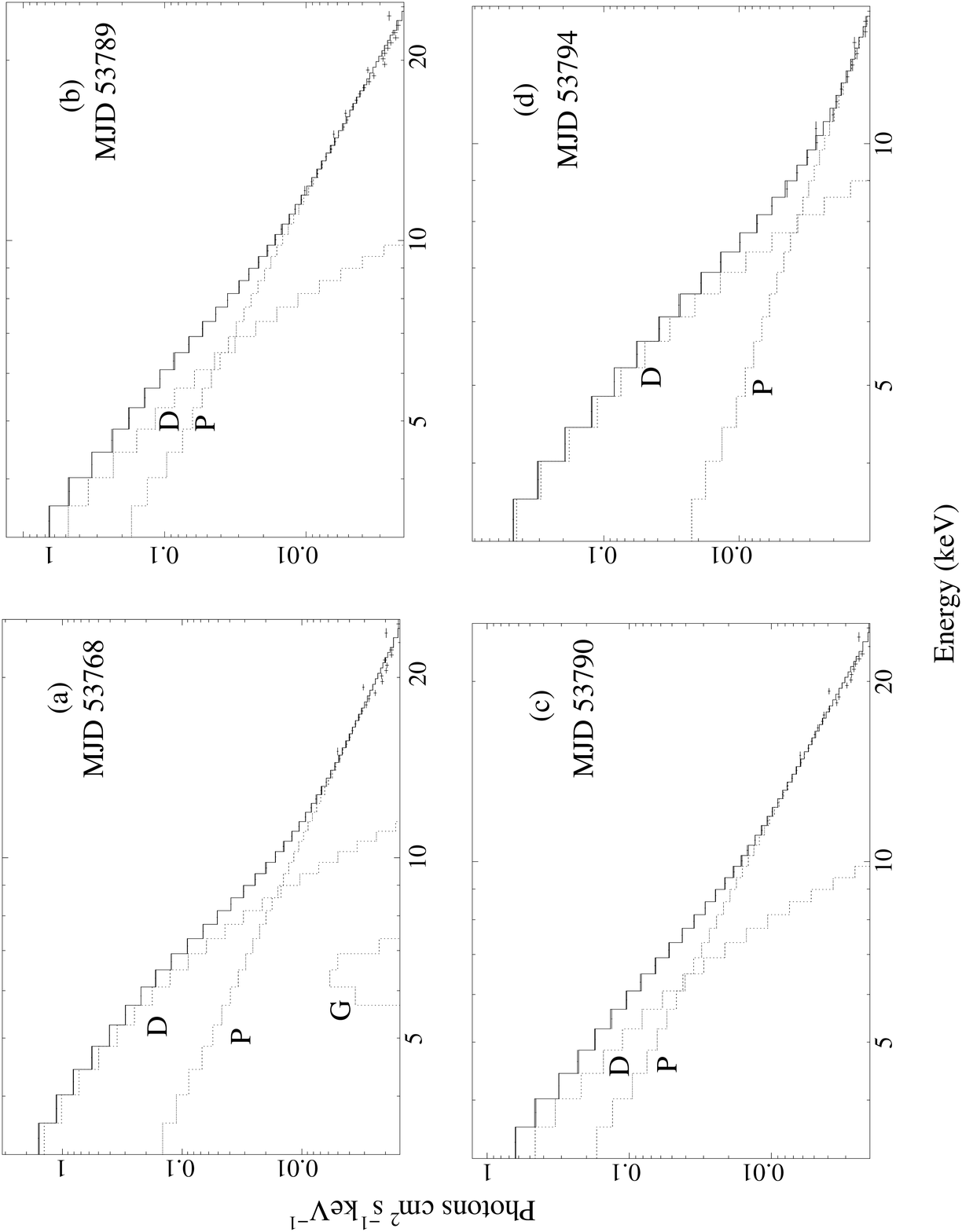}}
\caption{Some representative energy spectra with the unfolded models are shown in this figure. For details of the spectral model refer to page 5. Thick black lines indicates the total spectrum that is a sum of all the components. Symbol "D" denotes the disk black body component of the models used to fit the spectra, "P" indicates the power law component and gaussian line model is denoted by symbol "G". Fig (a) shows spectrum in the 3-25 keV for MJD 53768, (b) for MJD 53789 and (c) for MJD 53790 in which the QPOs are detected in the entire 3-25 keV energy range. Panel (d) shows spectrum in the 3-15 keV for MJD $53794$ in which no QPO is detected.}
\label{fig:fig8}
\end{figure}

The analyzed epochs included (a) MJD 53768, 53789, 53790 and 53790.6 that show presence of the QPOs in the entire 2-25 keV energy region, (b) MJD 53790.2 and 53790.3 in which broad QPO peaks in the 8-14 keV band are present but not in the 2-8 keV interval (c) MJD 53778, 53780, 53786 and 53790.5 data with the QPO detection only in the 2-8 keV band but not at $>$ 8 keV (d) MJD 53766 and 53791 data with the QPOs in the 2-14 keV interval but not at $>$ 15 keV (e) MJD 53764 and 53775 when no QPOs are detected (f) MJD 53794 and 53797 when the source intensity has declined and the QPOs are not present. 
In Table \ref{table2.tab} we have compiled the derived values of the temperature (T$_{in}$) of the disk black body component, photon spectral index ($\alpha$) of the hard component as well as the flux values of the thermal, and the power law components. Ratio of the power law flux to the thermal flux is also computed and shown in Table \ref{table2.tab}. Following points are to be noted from the table: (I) As expected for the black hole binaries the thermal component is dominant in the initial part of the outburst lasting for about first 25 days. This is obvious from the values of flux ratio that lies in 0.06 to 0.27 range. (II) After 1/e decay of the intensity, the thermal component declined substantially and the thermal and the power law fluxes became comparable. (III) When the intensity declined  further (after MJD 53791) the power law flux declines and the thermal flux dominates. Note that the power law spectral index lies in a narrow region of 2.1-2.3 for all the observations selected for the spectral studies. These values are comparable within the errors of the photon index values estimated by Gierlinski, Done and Page (2008) for some of the observations. The temperature of the disk is $\approx$ 1 keV and constant during the first 30 days but there is indication of cooling of the disk as shown by kT $\approx$ 0.8 with further decline of the intensity in the last two observations. A few representative energy spectra are shown in Fig \ref{fig:fig8}. The systematic errors for all the fits are within 3\%.


\section{Discussion}
\label{Discussion section}
The LFQPOs occur most frequently when the power law flux is the dominating component in the energy spectrum. Some times they are also present in the high luminosity state with the presence of a hard component. From table \ref{table1.tab} and \ref{table2.tab} it will be noticed that in all the cases of the detection of LFQPOs from XTE J1817-330, except the observations of MJD 53764 and 53775 in which no QPOs are detected, the ratio of the power law flux to the thermal disk flux lies in 0.20 to 1.13 range consistent with its occurrence only in the states with a significant power law component. Also note that the LFQPOs have significant coherence (Q $=$ $\nu$/$\delta\nu $) with the Q in range of 3-11 and their rms amplitude vary from a few percent to as high as 13\%.\\

\begin{figure}
\resizebox{\hsize}{!}{\includegraphics*[width=8cm, angle=-90]{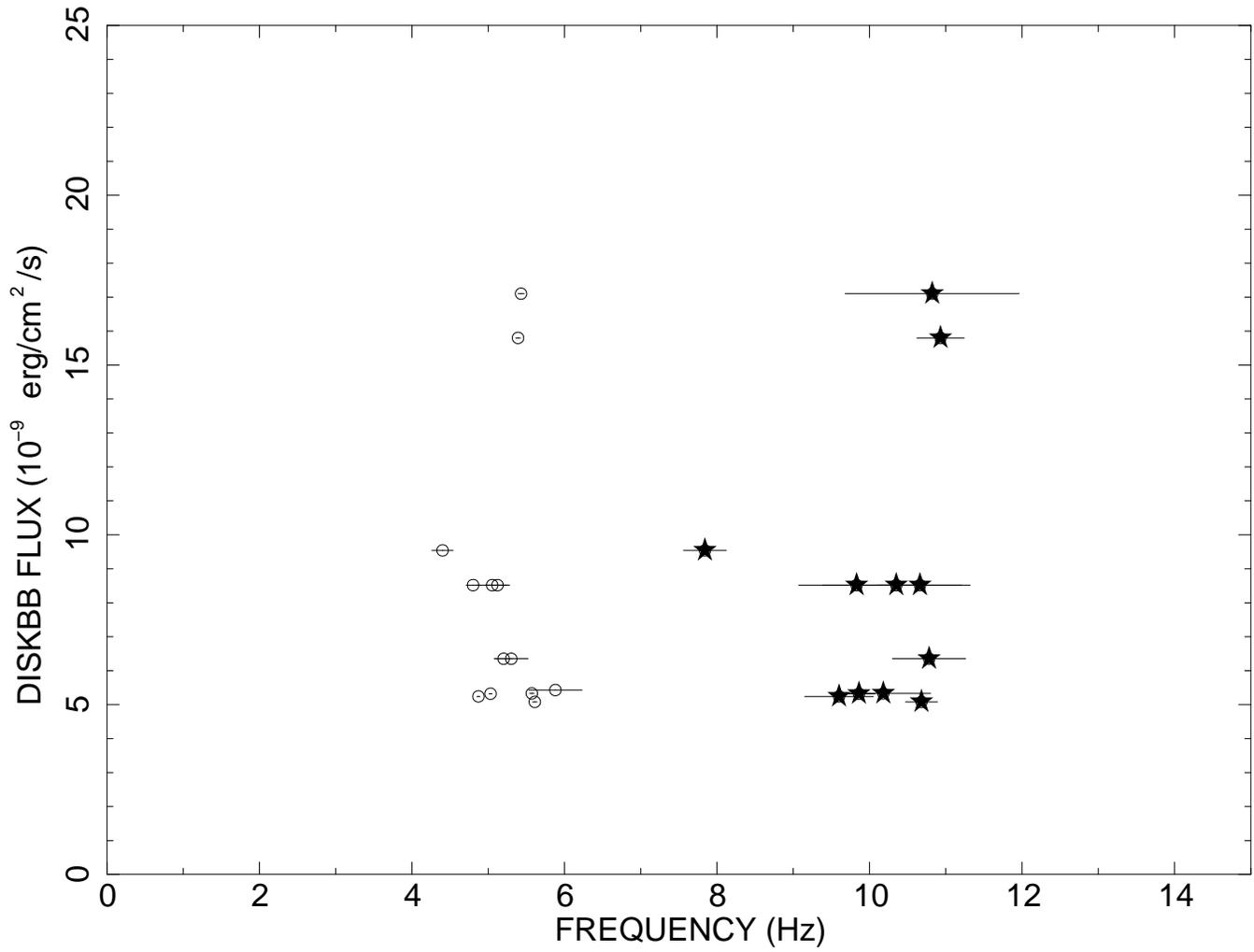}}
\caption{Variations of the QPO fundamental frequency in the 2-8 keV band (open circle) and first harmonics frequency (filled stars) with the flux of the disk black body component is shown for XTE J1817-330 during its 2006 outburst.}
\label{fig:fig9}
\end{figure}

Variation of the QPO frequency with the source intensity is another feature detected in some BHBs. The fundamental QPO frequency in XTE J1817-330 varies in a narrow band of 4.4-5.9 Hz. We have investigated the variation of the QPO frequency in the 2-8 keV band with the thermal disk component flux ( d$_{bb}$ ). This is shown in Fig \ref{fig:fig9} and a trend similar to that of Fig \ref{fig:fig7}(a) is seen here indicating that the frequency is correlated with the thermal disk component. As expected a clear 1:2 relationship of the QPO fundamental frequency and that of the first harmonic is seen.

All the characteristics of the LFQPOs reported by us in this paper from XTE J1817-330 are similar to those seen in the other black hole binaries and further strengthen the black hole nature of this source (Remillard et al. 2003).\\

Correlation of the properties of LFQPOs with the spectral parameters of the BHBs has been studied in detail for several sources (Muno, Remillard and Morgan 2001; Tomsick and Kaaret 2001; Remillard et al. 2003; Belloni, Psaltis and van der Klis 2002; Vignarca et al. 2003; Rossi, Homan and Belloni 2004). These studies show that the LFQPO characteristics are generally well correlated with the thermal disk and the power law components of the energy spectra. While the QPO frequency is closely correlated with the disk flux, the amplitude of the QPOs for the fundamental frequency is found to track the flux of the power law component (Remillard et al. 2003). In general the QPO amplitude is higher for the higher energy X-rays up to about 20 keV and then it tends to decrease at the higher energy. Most of the LFQPOs detected in the black hole binaries occur below 10 keV. In two of the BHBs namely GRS 1915+105 and XTE J1550-564 the QPOs have been reported above 20 keV (Trudolyubov, Churazov \& Gilfanov 1999; Remillard et al. 2003). In some cases the QPO detection is claimed in a broad spectral band of 2-60 keV but since no breakdown of QPO properties is given in the different energy intervals say below 10 keV and above 10 keV, it is not obvious whether the QPOs have indeed been detected in hard X-rays. The enigmatic source GRS 1915+105 is the only black hole binary in which the QPOs in 0.8-3.0 Hz have been reported at an energy up to 124 keV (Tomsick and Kaaret 2001). It is found that in this object, the amplitude of the fundamental frequency QPO increases with energy up to 29 keV and then decreases in 30-60 keV and 60-124 keV bands. Remillard et al. (2003) investigated the QPO characteristics in the transient XTE J1550-564 from the RXTE - PCA observations and detected LFQPOs in 2-13 keV and 13-30 keV energy channels making it only the second black hole binary in which the LFQPOs have been detected up to 30 keV. We have detected the QPOs from XTE J1817-330 in five of the observations up to an energy of $\sim$ 25 keV, making it only the third BHB showing unambiguous presence of the LFQPOs at higher energy. From Table \ref{table1.tab} it may be noticed that the amplitude of the fundamental QPOs is always higher in the 8-14 keV channel, being in 3.2-13.3 \% range, compared to the values of 1.7-7.0 \%  in the 2-8 keV channel. At still higher energy 15-25 keV, the QPO amplitude is comparable to or slightly higher than that in the 8-14 keV indicating that it has reached a plateau level.\\

A detailed study of the QPO centroid frequency, its coherence, amplitude, phase lag and their dependence on the photon energy, is of vital importance to understand the origin of the QPOs and pin point the emission process. The QPOs are believed to originate in the innermost region of the accretion disk and the most common models explain their generation to the modulation of the disk flux by the Keplerian motion of the localized hot regions termed as 'blobs'.  Lehr, Wagoner and Wilms (2000) have developed a model to compute by Monte Carlo simulations, the energy dependence of the QPO amplitude to probe the site of their origin in the accretion disk. They use two components with repeated Compton scattering to produce the high energy X-rays and assume a radial dependence of the disk temperature. They computed the energy dependence of the QPO amplitude for GRS 1915 + 105 and found it to be in agreement with the observation of Morgan and Remillard (1997). Thus they are able to localize the QPO origin in the inner disk. The amplitude of the QPOs will either increase or decrease with energy depending on the region of the disk in which the QPOs are produced and the temperature of the corona and its gradient. It is reasonable to assume that the QPO frequency is related to the dynamical time scale of the blobs and therefore, the LFQPOs observed by us in the 4.4-5.9 Hz from XTE J1817-330 will originate farther out in the accretion disk. In the outer region, the Compton scattering corona will be relatively cooler and the QPO amplitude will decrease with increasing energy. In our case we have detected a marginal increase in the amplitude of the QPOs at the higher energy in the two observations, a decrease in the amplitude in the other two observations and no change in amplitude in one case. This suggests that the site of QPO generation is itself dynamically varying in XTE J1817-330 due to variation in the X-ray luminosity which in turn depends on the accretion rate.\\
  
We have detected the QPOs in the 8-14 keV band but not in the 2-8 keV for the observations of MJD 53790.2 and 53790.3. This is similar to the detection of the QPOs at the high energy and its absence at the low energy in some observations from GRS 1915+105 (Chakraborti and Manickam 2000). This behavior of GRS 1915+105 was explained by Chakrabarti and Manickam (2000) on the basis of "on" and "off" (burst and quiescent) state of the source with the shock oscillation model. Further detailed studies of the LFQPOs at the higher energy are required to test the validity of the model. \\

\section{Acknowledgment}
\label{Acknowledgment section}
The authors thank NASA/GSFC based High Energy Astrophysics Science Archive Research Center for making RXTE-PCA data available on-line. We are extremely thankful to an anonymous referees for his critical and constructive comments and suggestions that vastly improved the content and presentation of this paper.

\appendix

\onecolumn
\begin{landscape}
\centering
\begin{table}
\small{
\caption{Summary of the characteristics of the QPO in the three energy bands. MJD 53766 corresponds to the date 2006-01-31.} 
\label{table1.tab}
\begin{tabular}{cccccccccccccc}
\hline\hline
MJD&Duration&Energy&Power Law&\multicolumn{4}{c}{QPO Fundamental}&\multicolumn{4}{c}{First Harmonic} &Reduced\\
\cline {5-8}
\cline {9-12}
   &Of Observation&Range&Index&Frequency & Width &Quality Factor&RMS&Frequency & Width &Quality Factor&RMS&Chi Sq$^\dag$  \\
   &(sec)&(keV)& & (Hz)&(Hz) &&\%&(Hz)&(Hz) &   &\%&\\
\hline\\
53766(Q1)& 3947&2-8&-1.25&$5.43_{-0.04}^{+0.04}$ & $0.81_{-0.10}^{+0.09}$&6.7 &2.5&$10.82_{-0.58}^{+0.56}$ & $1.42_{-0.98}^{+0.86}$&7.6 &0.9 &1.7  \\[2pt]
        &&8-14&-0.56&$5.42_{-0.20}^{+0.12}$&$ 0.62_{-0.36}^{+1.59}$& 8.7& 3.9& $10.86_{-0.50}^{+0.53}$&$ 2.62_{-1.99}^{+2.09}$&4.2 &6.8 &0.9\\[2pt]
53768(Q2) &5274 &2-8&-1.28&$5.39_{-0.02}^{+0.03}$&$ 0.89_{-0.07}^{+0.06}$& 6.0&2.6& $10.93_{-0.16}^{+0.15}$&$ 1.19_{-0.36}^{+0.27}$&9.2&1.0 & 2.4\\[2pt]
      	&&8-14&-0.57&$5.55_{-0.11}^{+0.13}$&$ 0.74_{-0.36}^{+0.33}$&7.4& 3.2&$10.89_{-0.10}^{+0.10}$&$ 2.15_{-0.33}^{+0.28}$&5.1&7.4 & 1.0 \\[2pt]
      &&15-25&-0.32&$ 5.14_{-0.23}^{+0.19}$ &$ 0.64_{-0.55}^{+0.58}$& 8.1 & 3.9&$10.83_{-0.31}^{+0.26}$&$ 2.22_{-0.92}^{+0.70}$& 4.9&8.1 & 1.0\\[2pt]
53778(Q3) & 10329&2-8&-1.19 &$4.41_{-0.13}^{+ 0.14}$ & $1.11_{-0.52}^{+ 0.34}$& 3.9&2.0 & $7.84_{-0.22}^{+0.06}$&$ 0.30_{-0.59}^{+1.19}$&26.1&1.0 &1.7 \\[2pt]
53780(Q4) &13549 &2-8&-1.09&$4.80_{-0.08}^{+ 0.09}$ & $0.79_{-0.29}^{+0.24}$ &6.1&2.2 &$9.83_{-0.52}^{+0.24}$&$ 0.91_{-1.72}^{+0.62}$& 10.8 & 1.3&1.2 \\[2pt]
      &&2-8&-1.25&$5.12_{-0.08}^{+0.08}$&$1.10_{-0.20}^{+0.17}$ & 4.7&2.5& $10.35_{-0.43}^{+0.54}$&$ 1.82_{-1.11}^{+0.85}$&5.7 &1.4 &1.6  \\[2pt]
      &&2-8&-1.87&$5.05_{-0.10}^{+0.10}$&$1.31_{-0.25}^{+0.21}$ & 3.8&2.5 & $10.66_{-0.31}^{+0.24}$&$ 0.67_{-0.97}^{+2.31}$& 15.8&1.0 &1.5\\[2pt]
53786(Q5)&8413 &2-8&-0.38&$5.29_{-0.13}^{+0.12}$ &$0.85_{-0.80}^{+0.35}$&6.2 & 2.5& -&-& -&-&1.3 \\[2pt]
	&&2-8&-1.11&$5.27_{-0.11}^{+0.11}$ &$1.53_{-0.38}^{+0.31}$ &3.5& 2.4&$10.78_{-0.24}^{+0.24}$&$ 1.42_{-0.66}^{+0.51}$ & 7.6&1.4&1.4 \\[2pt]
53789(Q6)&2769 &2-8&-1.09&$5.57_{-0.03}^{+0.03}$ & $0.94_{-0.06}^{+0.05}$&5.9 & 7.0&$10.18_{-0.30}^{+0.32}$&$ 2.92_{-1.00}^{+0.74}$&3.5&2.4& 1.5 \\[2pt]
     &&8-14&-2.42&$5.57_{-0.03}^{+0.03}$ & $0.95_{-0.07}^{+0.07}$ & 5.9&13.2 &$10.56_{-0.46}^{+0.45}$&$ 2.94_{-1.31}^{+0.93}$&3.6&5.2&1.2  \\[2pt]
     &&15-25&-1.82&$ 5.50_{-0.07}^{+0.07}$ &$1.01_{-0.16}^{+0.13}$&5.6 &11.5 &$10.97_{-2.31}^{+2.10}$&$ 2.96_{-3.29}^{+2.31}$ &3.7&4.5&0.9  \\[2pt]
53790(Q7)& 4132&2-8&-1.43&$5.61_{-0.03}^{+0.03}$ &$1.08_{-0.08}^{+0.07}$&5.2&6.7 &$10.68_{-0.37}^{+0.53}$ &$1.89_{-1.26}^{+0.76}$& 5.7&2.0 &1.6  \\[2pt]
	&&8-14&-2.48&$ 5.60_{-0.04}^{+0.04}$ &$ 1.06_{-0.11}^{+0.10}$&5.3&12.3 &$12.34_{-1.53}^{+1.25}$ &$5.27_{-6.8}^{+2.67}$ &2.3&6.0&0.9  \\[2pt]
      &&15-25&-2.11&$ 5.54_{-0.12}^{+0.13}$ &$ 1.38_{-0.35}^{+0.29}$&4.0 &12.0&-&-& -&-& 1.1 \\[2pt]
53790.2(Q8)& 3179&2-8&-1.09&NO QPO&-&-&-&-&-&-&-&1.0\\[2pt]
	&&8-14&-&$9.06_{-0.33}^{+0.32}$ &$2.97_{-1.95}^{+1.37}$&3.0&8.8 &-&-&-&-&1.2\\[2pt]
53790.3(Q9)&1967 &2-8&-1.12&NO QPO&-&-&-&-&-&-&-&1.1\\[2pt]
	& &8-14&-&$7.83_{-0.18}^{+0.16}$ &$2.03_{-0.90}^{+0.68}$&3.8&9.4 &-&-&-&- &1.2 \\[2pt]
        &&15-25&-&$7.97_{-0.37}^{+0.35}$ &$1.78_{-1.91}^{+1.28}$&4.5&9.6 &-&- &-&- &0.9\\[2pt]
53790.5(Q10)&3024 &2-8&-0.82&$5.88_{-0.37}^{+0.35}$ &$1.91_{-1.39}^{+0.77}$&3.1&1.7 &-&-&-&-&1.3 \\[2pt]
53790.6(Q11)& 3686&2-8&-1.16&$4.87_{-0.03}^{+0.02}$&$0.44_{-0.04}^{+0.04}$ &11.0&5.7&$9.60_{-0.23}^{+0.22}$&$1.20_{-0.69}^{+0.54}$&8.0&1.9 &1.6 \\[2pt]
       & &8-14& -1.29&$4.85_{-0.03}^{+0.02}$ &$ 0.44_{-0.05}^{+0.06}$&11.1 &12.9&$9.44_{-0.35}^{+0.37}$&$0.43_{-2.23}^{+2.23}$ & 22.1& 2.3& 1.0 \\[2pt]
       & &15-25&-0.42&$ 4.91_{-0.09}^{+0.09}$ & $ 0.55_{-0.13}^{+0.16}$&9.0&12.6 &-&-&-& -&1.7  \\[2pt]
53791(Q12) & 3564&2-8&-1.34 &$5.03_{-0.02}^{+0.02}$ &$ 0.58_{-0.03}^{+0.03}$&8.8&6.4&$9.86_{-0.10}^{+0.11}$&$1.26_{-0.25}^{+0.20}$& 7.8&2.4 & 1.6\\[2pt]
    & &8-14& -2.16&$ 5.01_{-0.02}^{+0.02}$ & $0.59_{-0.04}^{+0.04}$& 8.5&13.3 &$9.73_{-0.26}^{+0.27}$&$1.89_{-1.26}^{+0.72}$&5.2&4.9&1.7\\[8pt]
\hline
\end{tabular}
{\small \\[4pt]
$^\dag$ : Reduced chi square refer to the goodness of fit of the entire spectrum.\\ }
}
\end{table}
\end{landscape}
\twocolumn

\onecolumn
\begin{landscape}
\centering
\begin{table}
\scriptsize{
\caption{Summary of the spectral fit parameter for the selected observations of XTE J1817-330. MJD 53764 corresponds to the date 2006-01-29.}
\label{table2.tab}
\begin{tabular}{cccccccccccc}
\hline\hline
MJD   &DURATION OF & &power law&&           &diskbb&&Reduced&Energy&Total&Flux  \\
   & OBSERVATION&        &&&      &&&Chi Square&Range&Flux&Ratio$^{(1)}$  \\
\cline {3-8}
 &&Photon Index&Normalization&Flux&Tin&Normalization&Flux&&&  3-25 keV &        \\
   & (sec)& & &(ergs/cm$^2$/s) 10$^{-8}$ &(keV)&$\times$ 10$^{3}$ &(ergs/cm$^2$/s) 10$^{-8}$ & &(keV)&(ergs/cm$^2$/s 10$^{-8}$)&\\
\hline\\
53764	& 3117 &2.17($\pm0.24$)&0.54($\pm0.34$)&0.12	&0.99($\pm0.01$)&3.13($\pm0.41$)&1.59	&0.6&3-15&1.72&0.08	\\[6pt]
53766(Q1)& 3947 &2.31($\pm0.04$)&2.63($\pm0.34$)&0.45	&1.04($\pm0.01$)&2.52($\pm0.14$)&1.71	&1.1&3-20&2.16&0.27	\\[6pt]
53768(Q2)& 5274 &2.34($\pm0.04$)&2.57($\pm0.34$)&0.42	&1.03($\pm0.01$) &2.45($\pm0.21$)&1.58	&1.9&3-25&2.01&0.27	\\[6pt]
53775	& 3958&2.09($\pm0.27$)&0.25($\pm0.19$)&0.07	&0.95($\pm0.007$)&2.94($\pm0.12$)&1.21  &1.2&3-20&1.27&0.06	\\[6pt]
53778(Q3)& 10329 &2.08($\pm0.06$)&0.68($\pm0.18$)&0.19	&0.95($\pm0.01$)&2.45($\pm0.21$)&0.95   &1.1&3-20&1.14&0.20	\\[6pt]
53780(Q4)& 13549 &2.06($\pm0.06$)&0.61($\pm0.12$)&0.18	&0.94($\pm0.01$)&2.29($\pm0.18$)&0.85   &1.2&3-18&1.03&0.21	\\[6pt]
53786(Q5)& 8413&2.03($\pm0.06$)&0.48($\pm0.08$)&0.15	&0.91($\pm0.01$)&2.01($\pm0.16$)&0.63   &1.3&3-18&0.79&0.24	\\[6pt]
53789(Q6)& 2769&2.27($\pm0.03$)&2.79($\pm0.27$)&0.51	&1.05($\pm0.02$)&0.73($\pm0.87$)&0.53	&1.4&3-25&1.05&0.97	\\[6pt]
53790(Q7)& 4132&2.26($\pm0.03$)&2.74($\pm0.24$)&0.51	&1.05($\pm0.02$)&0.72($\pm0.88$)&0.51	&1.8&3-25&1.02&1.01	\\[6pt]
53790.2(Q8)& 3179 &2.13($\pm0.05$)&1.80($\pm0.26$)&0.44	&1.07($\pm0.02$)&0.59($\pm0.74$)&0.45	&1.2&3-20&0.89&0.99	\\[6pt]
53790.3(Q9)&1967 &2.15($\pm0.03$)&2.01($\pm0.21$)&0.47	&1.07($\pm0.02$)&0.53($\pm0.70$)&0.42	&1.3&3-25&0.89&1.13	\\[6pt]
53790.5(Q10)& 3024 &2.17($\pm0.04$)&1.58($\pm0.20$)&0.35&1.00($\pm0.02$)&0.94($\pm0.13$)&0.54	&1.5&3-25&0.89&0.65	\\[6pt]
53790.6(Q11)& 3686&2.26($\pm0.04$)&2.10($\pm0.27$)&0.39	&1.03($\pm0.02$)&1.03($\pm0.89$)&0.52	&1.3&3-20&0.91&0.75	\\[6pt]
53791(Q12)&3564&2.22($\pm0.05$)&2.08($\pm0.29$)&0.42	&1.04($\pm0.02$)&0.81($\pm0.93$)&0.53	&1.2&3-20&0.95&0.80	\\[6pt]
53794&3345 &2.08($\pm0.12$)&0.29($\pm0.09$)&0.07	&0.84($\pm0.01$)&2.14($\pm0.21$)&0.39 &1.1&3-15&0.47&0.20	\\[6pt]
53797&5539 &2.26($\pm0.10$)&0.39($\pm0.10$)&0.07	&0.82($\pm0.01$)&2.31($\pm0.18$)&0.35 &1.6&3-15&0.42&0.21	\\[6pt]
\hline
\end{tabular}
{\small \\[6pt]
$^{(1)}$:Ratio of power law flux to diskbb flux\\  
\\}
}
\end{table}
\end{landscape}
\onecolumn
\onecolumn

\onecolumn


\end{document}